\documentclass[fp]{jpsj3}
\usepackage{txfonts}

\title{Superconducting Fluctuation investigated by THz Conductivity of La$_{2-x}$Sr$_x$CuO$_4$ Thin Films}

\author{\name{Daisuke \surname{Nakamura}}\thanks{present address: dnakamura@issp.u-tokyo.ac.jp}, \name{Yoshinori \surname{Imai}}, and \name{Atsutaka \surname{Maeda}} \\
 $^{1}$\name{Ichiro \surname{Tsukada}}
}
\inst{\address{Department of Basic Science, the University of Tokyo, Meguro, Tokyo, 153-8902, Japan} \\
$^{1}$Central Research Institute of Electric Power Industry, Yokosuka, Kanagawa 240-0196, Japan
}
\abst{Frequency-dependent terahertz conductivities of La$_{2-x}$Sr$_x$CuO$_4$ thin films with various carrier concentrations were investigated.
The imaginary part of the complex conductivity considerably increased from far above a zero-resistance superconducting transition temperature, $T_\text{c}^\text{zero}$, because of the existence of the fluctuating superfluid density with a short lifetime.
The onset temperature of the superconducting fluctuation is at most $\sim 2T_\text{c}^\text{zero}$ for underdoped samples, which is consistent with the previously reported analysis of microwave conductivity.
The superconducting fluctuation was not enhanced under a 0.5 T magnetic field.
We also found that the temperature dependence of the superconducting fluctuation was sensitive to the carrier concentration of La$_{2-x}$Sr$_x$CuO$_4$, which reflects the difference in the nature of the critical dynamics near the superconducting transition temperature.
Our results suggest that the onset temperature of the Nernst signal is not related to the superconducting fluctuation we argued in this paper.}

\kword{cuprate superconductor, THz conductivity, superconducting fluctuation}

\begin{document}
\maketitle

\section{Introduction}

The physical properties of cuprate high-temperature superconductors (HTSC) are variable in the electronic phase diagram.
The superconducting transition temperature $T_\text{c}$ forms a bell-shape as a function of the carrier concentration $x$.
For a typical HTSC such as La$_{2-x}$Sr$_x$CuO$_4$ (LSCO), $T_\text{c}$ reaches a maximum value at $x \sim$ 0.15 (optimal doping).
In the underdoped and overdoped regions, different mechanisms are considered to control various properties in even the normal state.\cite{Timusk1999RPP}
For example, angle-resolved photoemmision (ARPES) measurements report a "pseudogap" in the normal state of HTSC, and the onset temperature of the pseudogap strongly depends on $x$.\cite{Hashimoto2007PRB, Yoshida2009PRL}
In particular, the pseudogap region of underdoped materials is much wider in temperature than that of overdoped materials, and the origin of the pseudogap is still under debate.\cite{Damascelli2003RMP}

Among various techniques and subjects, superconducting fluctuation is one of the issues most extensively investigated,\cite{TinkhamBook} to completely understand the mechanism underlying HTSC.
Theoretically, the thermal fluctuation of the order parameter of superconductivity causes the Gaussian superconducting fluctuation.
As a result, paraconductivity\cite{Aslamasov1968PLA, Maki1968RTP, Thompson1970PRB} is added to the normal-state conductivity.
For conventional low-temperature superconductors, the fluctuation effect above $T_\text{c}$ is well explained by this paraconductivity.\cite{Skocpol1975RPP}
In contrast, for HTSC, the critical fluctuation becomes dominant near $T_\text{c}$ because of the short coherence length, small superfluid density, quasi-two-dimensionality of the crystal structure, and high temperatures characteristic of HTSC.
In general, superconducting fluctuation can be described using the finite superfluid density $n_\text{s}$ with a short lifetime, which can be written using the amplitude of the order parameter ($\psi = |\psi|e^{i\theta}$) of the superconductor as $n_\text{s} /2 = |\psi|^2$.

There have been several experimental approaches to investigate the nature of the superconducting fluctuation in HTSC.
Because the imaginary part of the complex conductivity $\sigma_2$ in the superconducting state is directly related to $n_\text{s}$ as $\sigma_2 = n_\text{s}e^2 / m\omega$ ($e$ is the elementary charge, $m$ is the mass of an electron, and $\omega$ is the angular frequency), investigating the complex conductivity ($\tilde \sigma \equiv \sigma_1 +i \sigma_2$) at high frequencies is useful for the discussing $n_\text{s}$ with a short time scale.
Up to the present, there are many studies of $\tilde \sigma (\omega)$ at microwave frequencies in HTSCs; for example, HgBa$_2$CuO$_y$ (HBCO),\cite{Grbic2009PRB} YBa$_2$Cu$_3$O$_y$ (YBCO),\cite{Grbic2011PRB, Hosseini1999PRB} and LSCO.\cite{Kitano2006PRB, Ohashi2009PRB}
In particular, we previously investigated the frequency-dependent microwave (0.045-10 GHz) complex conductivity  of LSCO thin films with various carrier concentrations\cite{Kitano2006PRB, Ohashi2009PRB} and suggested that superconducting fluctuation becomes dominant only near $T_\text{c}$ (below $T \sim 2 T_\text{c}$ at most for the underdoped region) in the electronic phase diagram.
Because the upper limit of $T_\text{c}$ obtained by the microwave study was determined by extrapolating the microwave data to the infinite frequency ($\omega \to \infty$) using an existing theory of superconductivity fluctuation,\cite{Kitano2006PRB} a conductivity measurement at much higher frequencies is more preferable for the direct determination of the onset temperature of superconducting fluctuation.

In addition, in the microwave regime, we also found that optimally-doped and overdoped LSCO films have smaller regions of superconducting fluctuation than underdoped samples.
This interesting but puzzling doping dependence of superconducting fluctuation was concluded by the rapid change of the critical indices of superconducting fluctuation as a function of doping level, which is derived from dynamic scaling analysis of the complex conductivity.\cite{Ohashi2009PRB}
For the underdoped ($x \leq 0.14$) LSCO, two-dimensional $XY$ critical dynamics is observed, and the phase stiffness temperature $T_\theta$ introduced by Berezinskii-Kosterlitz-Thouless (BKT)\cite{Berezinskii1970JETP, Kosterlitz1973JPC} defined as
\begin{equation}
 k_B T_\theta = \frac{\hbar}{e^2} \hbar \omega \sigma_2 d_s
 \label{eq:BKT}
\end{equation}
($k_\text{B}$ is Bortzmann's constant, $\hbar$ is Planck's constant, and $d_s$ is the thickness of the two-dimensional superconductor) shows the characteristic temperature-dependent nature of the BKT transition.
Namely, $T_\theta$ shows a universal jump from zero to $(8/\pi) T_\text{BKT}$ at the BKT phase transition temperature, $T_\text{BKT}$.
In contrast, in the optimally-doped ($0.15 \leq x \leq  0.16$) region, three-dimensional $XY$ critical dynamics is observed.
This critical dynamics agrees with many published results for other HTSCs, YBCO\cite{Salamon1993PRB, Kamal1994PRL, Overend1994PRL, Pasler1998PRL} and Bi$_2$Sr$_2$CaCu$_2$O$_{8+\delta}$ (BSCCO),\cite{Osborn2003PRB} which indicates that there is common mechanism of the critical dynamics in optimally-doped HTSC.
Surprisingly, in the overdoped ($0.17 \leq x$) region, the superconducting fluctuation shows the behavior typical for two-dimensional superconductors again, but with critical indices different from any well-known type of phase transition, such as the $XY$-model, and Gaussian fluctuation $etc$.
We have argued that these behaviors are related to the phase diagram of cuprate superconductors with a "hidden" quantum critical point.\cite{Ohashi2009PRB}
It should be noted that the above mentioned changes in the critical indices as a function of doping take place very sharply.
Therefore, it will be very interesting to determine whether this characteristic doping dependence of the universality class is also observed at much higher frequencies.

In contrast to $\tilde \sigma (\omega)$ at high frequencies, the Nernst and diamagnetic signals appear far above $T_\text{c}$ (e.g., 130 K at $x$ = 0.10 and $\sim$ 100 K at optimally doped region), especially in the underdoped sample, which has been suggested to be the vortex-like precursor of the superconducting transition.\cite{Xu2000Nature,Wang2001PRB,Li2010PRB}
However, the onset temperature of the Nernst signal in other materials, such as the optimally doped BSCCO and YBCO, is close to that of LSCO ($\sim$ 105 K for YBCO, $\sim$ 125 K for BSCCO),\cite{Wang2006PRB,Li2010PRB} despite these materials having much higher $T_\text{c}$ than LSCO.
In addition, recent experiments on Nd-doped and Eu-doped LSCO have suggested that the Nernst signal in these materials comes from the stripe order.\cite{Cyr-Choiniere2009Nature}
Thus, the origin of the Nernst signal is still controversial in terms of the relevance to superconductivity.
We note that the Nernst and diamagnetism experiments are typically performed under a finite magnetic field, despite complex conductivity measurement usually performed without magnetic field.
Therefore, the Nernst signal and the microwave conductivity might be caused from different physical phenomena.
Since superconducting fluctuation is usually suppressed by the magnetic field, the characteristic Nernst and diamagnetism signals in the normal state are considered not to be related to the superconducting fluctuation we argued in this paper.

Based on this background, in this paper, we perform a systematic complex conductivity study of LSCO thin films with various doping levels in the THz frequency region between 0.2-2.0 THz, including the effect of a finite magnetic field.
Though the onset temperature of the superconducting fluctuation was already discussed using THz spectroscopy in YBCO\cite{Nuss1991PRL} and BSCCO,\cite{Corson1999nature, Murakami2002JS, Orenstein2006AP} there is little description about the difference of the superconducting fluctuation nature along the entire doping range.
Therefore, investigating the difference of the nature of superfluid and quasiparticle from THz conductivity spectrum in HTSC samples with various carrier concentration is considered to be valuable for discussion of the superconducting fluctuation.
Therefore, we will focus on the following three aspects: (1) the onset temperature of superconducting fluctuation, (2) the doping dependence of superfluid and quasiparticle properties near $T_\text{c}$ and in the superconducting state, and (3) the magnetic field effect of the superconducting fluctuation.
Some of the results have already been published in brief papers.\cite{Nakamura2009M2S,Maeda2010PhysC}
It should be noted that, upon completion of this work, we noticed that results consistent with ours, also obtained by the THz technique, were published, concerning point (1).\cite{Bilbro2011NPhys}

\section{\label{sec:2}Experiments and Data Analysis}
\begin{center}
\begin{table}[t]
\centering
\caption{Specifications of LSCO thin films. The nominal Sr concentration ($x$), the thickness of the film ($d$), the $c$-axis length ($c_0$), the onset temperature of the superconducting transition ($T_\text{c}^{\text{onset}}$), the zero resistance temperature ($T_\text{c}^{\text{zero}}$), and the onset temperature of the superconducting fluctuation ($T_\text{fluct}$), are shown.}
\label{sample_table}
\begin{tabular}{cccccc}
\hline
\hline
$x$ & $d$ (nm) & $c_0$ (\AA) & $T_\text{c}^{\text{onset}}$ (K) & $T_\text{c}^{\text{zero}}$ (K) & $T_\text{fluct}$ (K) \\
\hline
0.07 & 120 & 13.240 & 22.1 & 18.5 & $\sim$ 35 \\
0.10 & 120 & 13.216 & 27.6 & 18.7 & $\sim$ 40-45 \\
0.12 & 160 & 13.229 & 32.9 & 27.9 & $\sim$ 50 \\
\hline
0.15 & 82 & 13.227 & 29.0 & 25.0 & $\sim$ 30 \\
0.16 & 80 & 13.242 & 36.9 & 32.0 & $\sim$ 40 \\
\hline
0.17 & 115 & 13.230 & 33.4 & 30.3 & $\sim$ 40 \\
0.225 & 100 & 13.229 & 16.1 & 10.9 & $\sim$ 15-20 \\
\hline
\hline
\end{tabular}
\end{table}
\end{center}

\begin{figure}[!t]
\begin{center}
\includegraphics[width=1.00\linewidth]{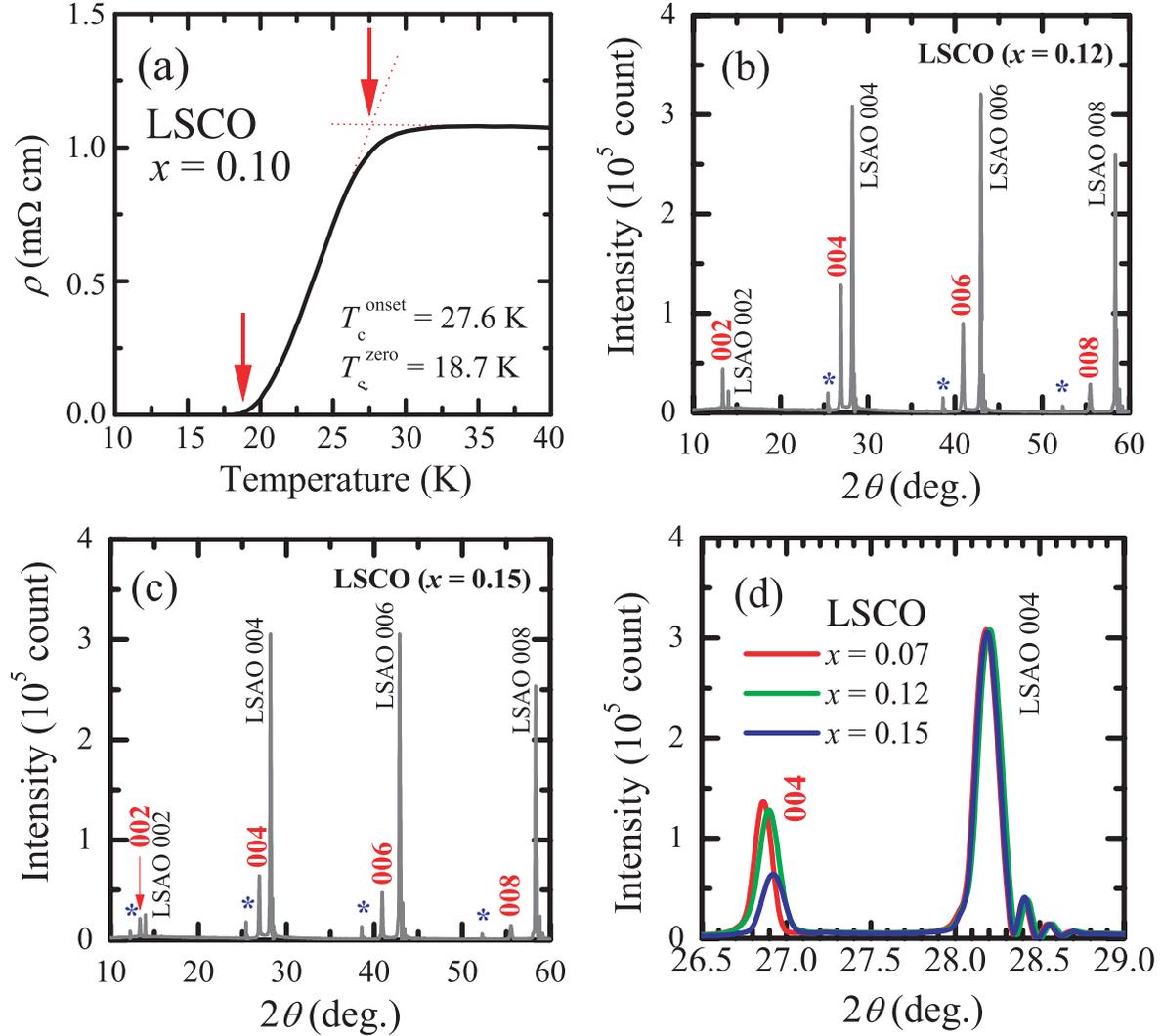}
\caption{ (Color online) (a) Temperature dependence of the dc resistivity of the underdoped LSCO ($x$ = 0.10) film near $T_\text{c}$. Down arrows indicate $T_\text{c}^\text{onset}$ and $T_\text{c}^\text{zero}$. 
(b)(c) X-ray diffraction spectrum of (b) the underdoped LSCO ($x$ = 0.12) and (c) the optimally-doped LSCO ($x$ = 0.15) films. 
Bold Miller indices indicates the 00$l$ peaks of LSCO films, and $\ast$ symbols indicate the LSAO peaks from Cu-K$\beta$ line.
(d) Comparison of X-ray diffraction spectrum near 004 peak in LSCO films with different carrier concentrations. }
\label{Fig:x010RT}
\end{center}
\end{figure}

\begin{figure*}[t]
\begin{center}
\includegraphics[width=1.0\linewidth]{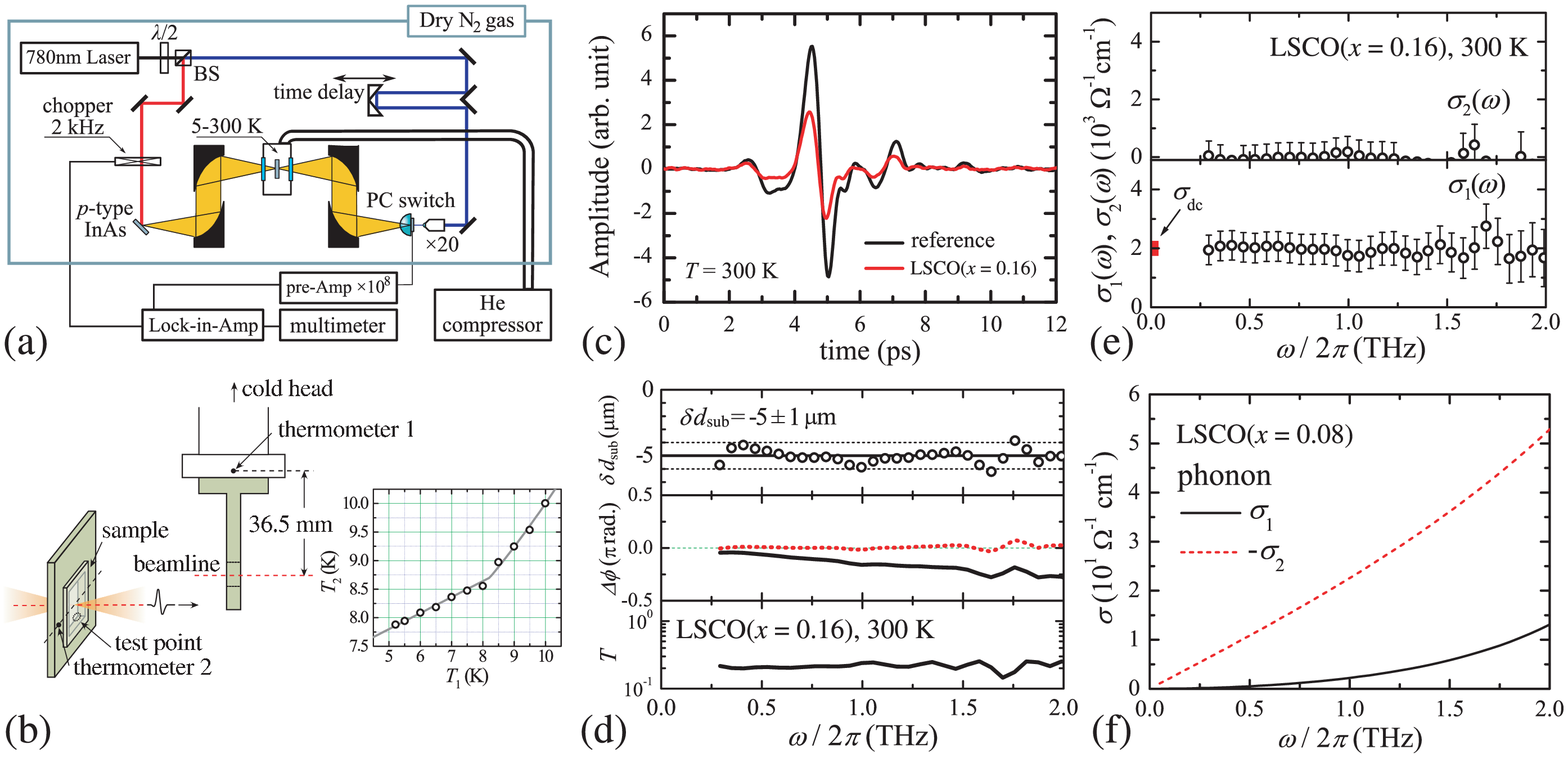}\\
\vspace{-0.3cm}
\caption{\label{Fig:Setup} (Color). 
(a) Schematic figure of the experimental setup. Details are described in the text. (b) Arrangement of the sample holder (gray region) and the thermometer in the cryostat. Right panel compares the temperature of the thermometer 1 ($T_1$) with that of the thermometer 2 ($T_2$). (c) Time-domain waveforms through the reference (LSAO) and the sample (optimally doped LSCO, $x$ = 0.16) at the room temperature. (d) Frequency dependence of the transmittance, $T$, the phase shift, $\Delta \phi$, and the difference of the substrate thickness, $\delta d_\text{sub}$, which are calculated from the data in Fig. \ref{Fig:Setup}(c). The red dotted line in the middle panel is $\Delta \phi$ after the calibration. (e) Frequency dependence of the complex conductivity, $\tilde \sigma (\omega)$, of LSCO ($x$ = 0.16) film at the room temperature. The lower panel is the real part, $\sigma_1$, and the upper panel is the imaginary part, $\sigma_2$. (f) $\tilde \sigma (\omega)$ by LSCO ($x$ = 0.08) phonon calculated from the data in Ref. 47. The solid line and the dotted line are $\sigma_1 (\omega)$ and $-\sigma_2 (\omega)$, respectively. 
\vspace{-0.5cm}}
\end{center}
\end{figure*}

We prepared $c$-axis oriented LSCO thin films with various doping levels ($x$ = 0.07-0.225) by the pulsed laser deposition (PLD) method on a 0.5 mm thick double-side polished LaSrAlO$_4$ (LSAO) substrate.
Details on the fabrication of films are described elsewhere.\cite{Lavrov2003PRB, Tsukada2004PRB, Tsukada2006PRB}
Especially in underdoped HTSC, the effect of the ordered state of holes and spins should be carefully considered in the normal state, because it often causes some anomaly in various physical properties and makes the contribution of the superconducting fluctuation obscure.
From this reason, we selected LSCO/LSAO film for an experimental target, because it shows high $T_\text{c}$ without the depression of $T_\text{c}$ at $x \sim 0.125$, caused by the stripe order of holes and spins.\cite{Tranquada1995Nature}
Table I shows the nominal Sr concentration ($x$), the thickness of the film ($d$), the lattice constant along the $c$-axis ($c_0$), the onset temperature of the superconducting transition ($T_\text{c}^\text{onset}$), the zero dc resistance temperature ($T_\text{c}^\text{zero}$), and the onset temperature of the superconducting fluctuation obtained in this work ($T_\text{fluct}$).
$T_\text{c}^\text{onset}$ and $T_\text{c}^\text{zero}$ were determined from the dc resistivity as shown in Fig. \ref{Fig:x010RT}(a) in the case of $x$ = 0.10 film (down arrows).
An X-ray diffraction measurement (Cu-K$\alpha$ line, diffraction method) was performed to obtain $c_0$.
The X-ray spectra of LSCO films are shown in Fig. \ref{Fig:x010RT}(b) for the underdoped LSCO ($x$ = 0.12) film and in Fig. \ref{Fig:x010RT}(c) for the optimally doped LSCO ($x$ = 0.15) film.
Bold Miller indices indicate the 00$l$ peaks of LSCO films, and $\ast$ symbols indicate the LSAO peaks from Cu-K$\beta$ line.
We also show comparison of X-ray diffraction spectrum near the 004 reflections in LSCO films with different carrier concentrations (Fig. \ref{Fig:x010RT}(d)).
From these figures, we confirmed that the peak width of the diffraction peaks of LSCO films are comparable to those of the LSAO substrates, supporting the crystallographic quality of our thin-film samples are sufficient for this study. 
The $c_0$ values of LSCO films in Table I are above 13.21 \AA , which indicates that all the films are compressed along the CuO$_2$ plane direction.\cite{Sato2000PRB}

An LSAO substrate was fixed to the holder with silver paste (DuPont 4922N) at its backside during deposition to maintain a homogeneous thermal distribution on the substrate.
Because residual silver paste directly affects the results of transmitted-type measurements, we dissolved the remaining silver paste in a solvent after deposition.
However, because of high substrate temperatures ($\sim$ 820-840 ${}^\circ$C) during the film growth, silver paste sometimes bakes on the reverse side of the substrate.
In such cases, we carefully removed the silver paste by polishing with lapping film.

The frequency-dependent (0.2-2.0 THz) complex conductivity was measured by transmitted-type THz time-domain spectroscopy (THz-TDS),\cite{Grischkowsky1990JOSAB} as schematically shown in Fig. \ref{Fig:Setup}(a).
For the emitter and detector of the THz pulse, we used Zn-doped InAs \cite{Gu2002JAP} and a dipole-type photoconductive switch (PC switch),\cite{Auston1984APL} respectively.
The radiated THz pulse from the emitter in air (yellow hatched region) is concentrated by Al-coated off-axis parabolic mirrors, so that it transmits through LSCO films before detected by PC switch.
The pump beam (red line) for the THz pulse emitter was modulated at 2 kHz for phase-locked detection.
Two time-domain waveforms of the transmitted THz-pulse through the sample (LSCO film on substrate) and reference (LSAO substrate only) were measured.
The measurement system is continuously purged with dry nitrogen gas to eliminate the effect of absorption by water vapor.
The typical value of room temperature fluctuation during the experiments is approximately $\pm$ 2.0 K.
In a Gifford-MacMahon refrigerator-type cryostat with quartz optical windows (Opticool, Oxford Instruments), we set the sample so that the electric field of the THz pulse was parallel to the CuO$_2$ plane of the LSCO films and hence, we measured in-plane conductivity in this work.
Figure \ref{Fig:Setup}(b) shows the arrangement of the sample holder (gray region) in the cryostat.
The sample is attached to the oxygen-free Cu sample holder carefully using Ag paste so as to leave the measurement area clean. 
Because the RhFe sensor (thermometer 1 in Fig. \ref{Fig:Setup}(b)) of the cryostat is separated from the working area of the sample by approximately 36.5 mm, the temperature of thermometer 1, $T_1$, is not exactly the same as that of the sample near the lowest temperature.
Therefore, we attached a Cernox 1050 thermometer (thermometer 2 in Fig. \ref{Fig:Setup}(b)) close to the sample and measured $T_1$ and the temperature of the thermometer 2, $T_2$, simultaneously.
The right panel of Fig. \ref{Fig:Setup}(b) shows the difference between $T_1$ and $T_2$, and we calibrated the temperature.
We also separately confirmed good thermal contact of LSCO films to the sample holder by attaching a Cernox 1050 thermometer just below the working area of the sample during the irradiation of the THz pulse (test point in Fig. \ref{Fig:Setup}(b)).
Therefore, we believe that local heating of the sample can be ignored.

Figure \ref{Fig:Setup}(c) shows the typical time-domain waveforms through the reference (LSAO) and sample (optimally doped LSCO, $x$ = 0.16) at room temperature.
We obtained the frequency-dependent transmittance, $T(\omega)$, and the phase shift, $\Delta\phi (\omega)$, by dividing the Fourier-transformed signal through the sample by that through the reference.
Solid lines in the lower panel and in the middle panel of Fig. \ref{Fig:Setup}(d) are $T(\omega)$ and $\Delta\phi (\omega)$, respectively.
The complex refractive index of the film, $\tilde n \equiv n + i\kappa$, can be obtained by fitting ($T(\omega), \Delta\phi (\omega)$) to the following analytical formula:

\begin{equation}
\begin{split}
 \sqrt{T(\omega)} \exp [-i\Delta \phi (\omega)]  & \\
 = \frac{2\tilde n (\tilde n_{\text{sub}}+1)}{(\tilde n + 1)(\tilde n + \tilde n_{\text{sub}})} & \frac{\exp \left[-i \frac{(\tilde n -1) d\omega}{c} \right]}{1-\frac{\tilde n - 1}{\tilde n + 1}\frac{\tilde n - \tilde n_{\text{sub}}}{\tilde n + \tilde n_{\text{sub}}} \exp \left[-i \frac{2\tilde n d\omega}{c} \right] } ,
\end{split}
\end{equation}
where $\tilde n_{\text{sub}}$ is the complex refractive index of the LSAO substrate, and $c$ is the speed of light.
The simplex method was utilized to iteratively fit the data.\cite{Nelder1965CJ}
Next, the complex dielectric function, $\tilde \varepsilon \equiv \varepsilon_1 - i\varepsilon_2 = \varepsilon_0 \tilde n^2$, where $\varepsilon_0$ is the permittivity of vacuum, and the complex conductivity, $\tilde \sigma \equiv \sigma_1 + i\sigma_2 = -i \omega (\tilde \varepsilon - \varepsilon_0)$, were calculated.
We separately measured $\tilde n_{\text{sub}}$ of the LSAO substrate as $n_{\text{sub}} \sim 4.2$ and $\kappa_{\text{sub}} \sim 0.01$ at 1.0 THz, which is consistent with the previously reported value.\cite{Shannon1992JSSC, Kamba1998PSSA}
With this method, $\tilde n (\omega)$ can be obtained without the Kramers-Kronig transformation, which requires a very broad spectral range to obtain the correct frequency-dependent complex conductivity, $\tilde \sigma (\omega)$.
Therefore, we can directly obtain $\tilde \sigma (\omega)$ value without the ambiguity induced by the Kramers-Kronig transformation.

In transmitted-type THz-TDS, an intrinsic difference in the substrate thickness, $\delta d_\text{sub}$, can cause a serious error in $\Delta \phi$.
To eliminate this effect, we corrected the obtained phase shift in the following manner.
Because $n_{\text{sub}} \gg \kappa_{\text{sub}}$, $\delta d_\text{sub}$ does not substantially affect the transmittance.
If we suppose that $\delta d_\text{sub}$ = 20 $\mu $m, the error in the transmittance, $\delta T$, is roughly estimated as $\delta T = 1- \exp [-2\omega \kappa_{\text{sub}} \delta d_\text{sub} / c] = 0.008$ at 1.0 THz for the LSAO substrate, which is negligibly small. 
Therefore, the experimentally obtained phase shift, $\Delta \phi_{\text{exp}}$, alone needs to be calibrated.
For this calibration, we assumed that $n = \kappa $ at the highest experimental temperature (250-300 K), indicating that $\sigma_2 \sim \omega \varepsilon_0 (\kappa^2 - n^2) = 0$.
This corresponds to the low-frequency response of Drude metal and is widely used to analyse the microwave conductivity of metallic materials.\cite{Maeda2005JPCM}
Therefore, using experimentally obtained $T$ and with the assumption $n-\kappa = 0$, we calculated $\Delta\phi_{n=\kappa}$ and $n+\kappa$ by Eq. (2) at each frequencies, where $\Delta\phi_{n=\kappa}$ is the phase shift in the case of $n = \kappa$.
Then we extracted $\delta d_\text{sub} = ( \Delta \phi_{\text{exp}} - \Delta\phi_{n=\kappa} ) c / \omega n_{\text{sub}}$ (about -5 $\mu$m in the case of $x$ = 0.16 film, upper panel of Fig. \ref{Fig:Setup}(d)), which is close to the value roughly measured by the micrometer.
Next, we subtracted $\Delta \phi_{\text{exp}} - \Delta\phi_{n=\kappa}$ obtained at the highest experimental temperature from $\Delta \phi_{\text{exp}}$ of the whole temperature region.
This calibration of $\Delta \phi (T)$ is valid if $\delta d_\text{sub}$ does not change with the temperature.
The red dotted line in the middle panel of Fig. \ref{Fig:Setup}(d) is the calibrated phase shift, $\Delta \phi_{\text{calib}}$.
After this calibration, we calculated $\tilde \sigma (\omega)$ of the film by fitting ($T(\omega), \Delta \phi_{\text{calib}} (\omega)$) to Eq. (2).
Figure \ref{Fig:Setup}(e) shows the $\tilde \sigma (\omega)$ of LSCO ($x = 0.16$) film.
The real part ($\sigma_1$, the lower panel) is almost equal to the dc conductivity ($\sigma_\text{dc}$, square symbol) and is independent of frequency, and the imaginary part ($\sigma_2$, the upper panel) is almost zero.

We confirmed the validity of the assumption $\sigma_2 \sim 0$ at the highest temperature of the measurement as follows.
Even in the normal state, a Drude-type conductivity spectrum ($\sigma_1 = \sigma_\text{dc}/(1+(\omega\tau)^2), \sigma_2 = \sigma_\text{dc}\omega\tau/(1+(\omega\tau)^2)$) leads to a finite $\sigma_2$ when the carrier lifetime ($\tau$) is large.
Suzuki $et\ al.$ reported $\tau$ of LSCO with various doping levels using optical reflectivity measurements at room temperature.\cite{Suzuki1989PRB}
According to the authors, $\tau \sim 0.9 \times 10^{-14}$ s for the optimally doped ($x$ = 0.15) sample, leading to $\omega \tau = 0.057$ at 1.0 THz.
If we assume a dc resistivity of optimally doped ($x$ = 0.15) LSCO at room temperature of $\rho_\text{dc} = 3.6 \times 10^{-4} \ \Omega$ cm,\cite{Ando2004PRLv2} we obtain $\sigma_2 \sim 158 \ \Omega^{-1} \text{cm}^{-1}$.
This value is less than the experimental error of our measurement.

In addition, at higher frequencies (for example, in the optical region), the conductivity spectrum can not be described only by that of the Drude-type carrier.\cite{Tanner1992Ginsberg, Gao1993PRB, Tajima2005PRB}
Using the well-known two-component approach, the measured dielectric function can be described by the following additive superpositions: $\tilde \varepsilon (\omega) = \tilde \varepsilon_{\text{Drude}} (\omega) + \tilde \varepsilon_{\text{Lorentz}} (\omega) + \varepsilon_{\infty}$, where $\tilde \varepsilon_{\text{Drude}}$ is the contribution of the Drude-like free carrier, $\tilde \varepsilon_{\text{Lorentz}}$ is that of the Lorentz-type oscillator, and $\varepsilon_{\infty}$ is a constant value in the high-frequency limit.
In HTSC, some types of bound carriers (e.g., phonon, mid-infrared absorption, charge transfer excitation, and interband transition) contribute to the measured complex conductivity as the Lorentz oscillator.\cite{Gao1993PRB}
We expect that these contributions can be mostly neglected in our experimental frequency region.
To check this, we explicitly estimate the contribution of the vibration mode of phonon as an example, which has the lowest frequency among the bound carrier contributions to $\tilde \sigma (\omega)$.
We calculated the $\tilde \sigma (\omega)$ of phonon in the THz region (shown in Fig. \ref{Fig:Setup}(f)) using the previously reported phonon parameter for the LSCO ($x$ = 0.08) single crystal.\cite{Collins1989PRB}
The solid and dotted lines in Fig. \ref{Fig:Setup}(f) are $\sigma_1 (\omega)$ and $-\sigma_2 (\omega)$, respectively.
Though there is some temperature dependence \cite{Gao1993PRB} and doping dependence \cite{Herr1987PRB} of the phonon parameter in LSCO, the $\tilde \sigma (\omega)$ of phonon is still far smaller than that of the free carrier in the THz range.
Therefore, we assume that the contribution of bound carriers can be ignored in the measured conductivity data, under the overall temperature and doping ranges. 
From these verification, we judged that the assumption of $\sigma_2 \sim 0$ is valid at the highest measured temperature.

\begin{figure}[!t]
\begin{center}
\begin{minipage}{0.99\linewidth}
\includegraphics[width=0.99\linewidth]{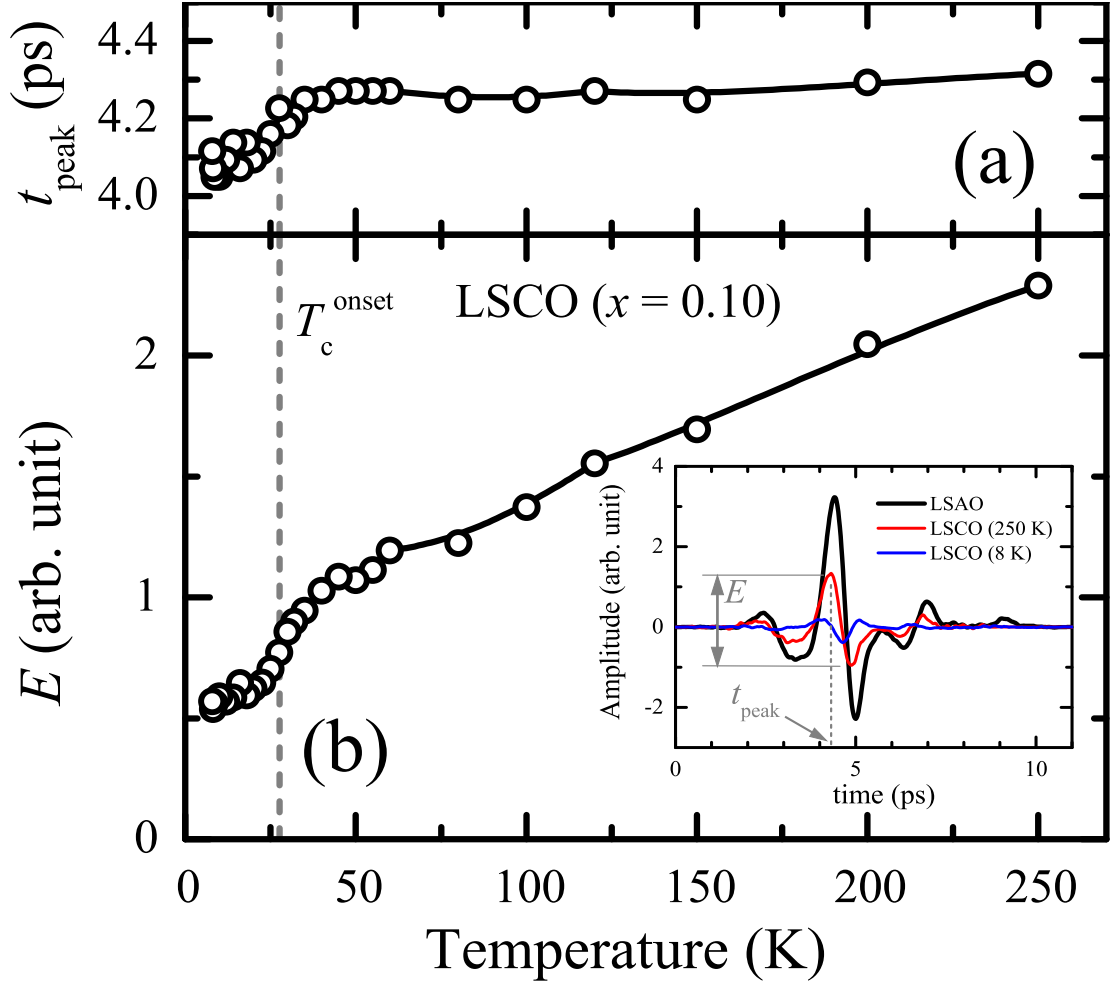}
\end{minipage}
\caption{\label{fig:x010WF} (Color online) Temperature dependence of amplitude ($E$, lower panel) and peak time ($t_\text{peak}$, upper panel), where $E$ reaches a maximum value for the underdoped LSCO ($x$ = 0.10) film. The inset in the lower panel shows time-domain waveforms of the transmitted THz pulse through the LSAO substrate (black: 250 K) and LSCO ($x$ = 0.10) film (red: 250 K, blue: 8 K).}
\end{center}
\end{figure}

\section{\label{sec:level3}Results and discussions}

First, we will show the results for different doping regions separately, and the Sr doping dependence will subsequently be discussed.

\subsection{\label{sec:level3-1}underdoped region: 0.07 $\leq x \leq $ 0.12}

The underdoped LSCO ($x$ = 0.10) film shows that $T_\text{c}^\text{onset}$ = 27.6 K and $T_\text{c}^{\text{zero}}$ = 18.7 K.
Figure \ref{fig:x010WF} shows the temperature dependence of the amplitude ($E$, lower panel) and the peak time ($t_\text{peak}$, upper panel), where $E$ reaches a maximum value, of the transmitted THz pulse through LSCO film.
The inset of Fig. \ref{fig:x010WF} shows time-domain waveforms of the transmitted THz pulse through the LSAO substrate (black: 250 K) and LSCO film (red: 250 K, blue: 8 K).
$E$ decreases with temperature in the normal state, which is due to the reduction of dc resistivity.
In the superconducting state, the amplitude of the transmitted THz pulse is suppressed further.
In addition, $t_\text{peak}$ starts to decrease rapidly from $T \sim$ 40 K, which causes an increase in the phase shift and $\sigma_2$.
These behaviors are what we expect for superconductors.
Indeed, the above features are similar to a previous report on the THz conductivity of Bi-cuprate HTSC, for example.\cite{Murakami2002JS}

\begin{figure}[!t]
\begin{center}
\includegraphics[width=0.8\linewidth]{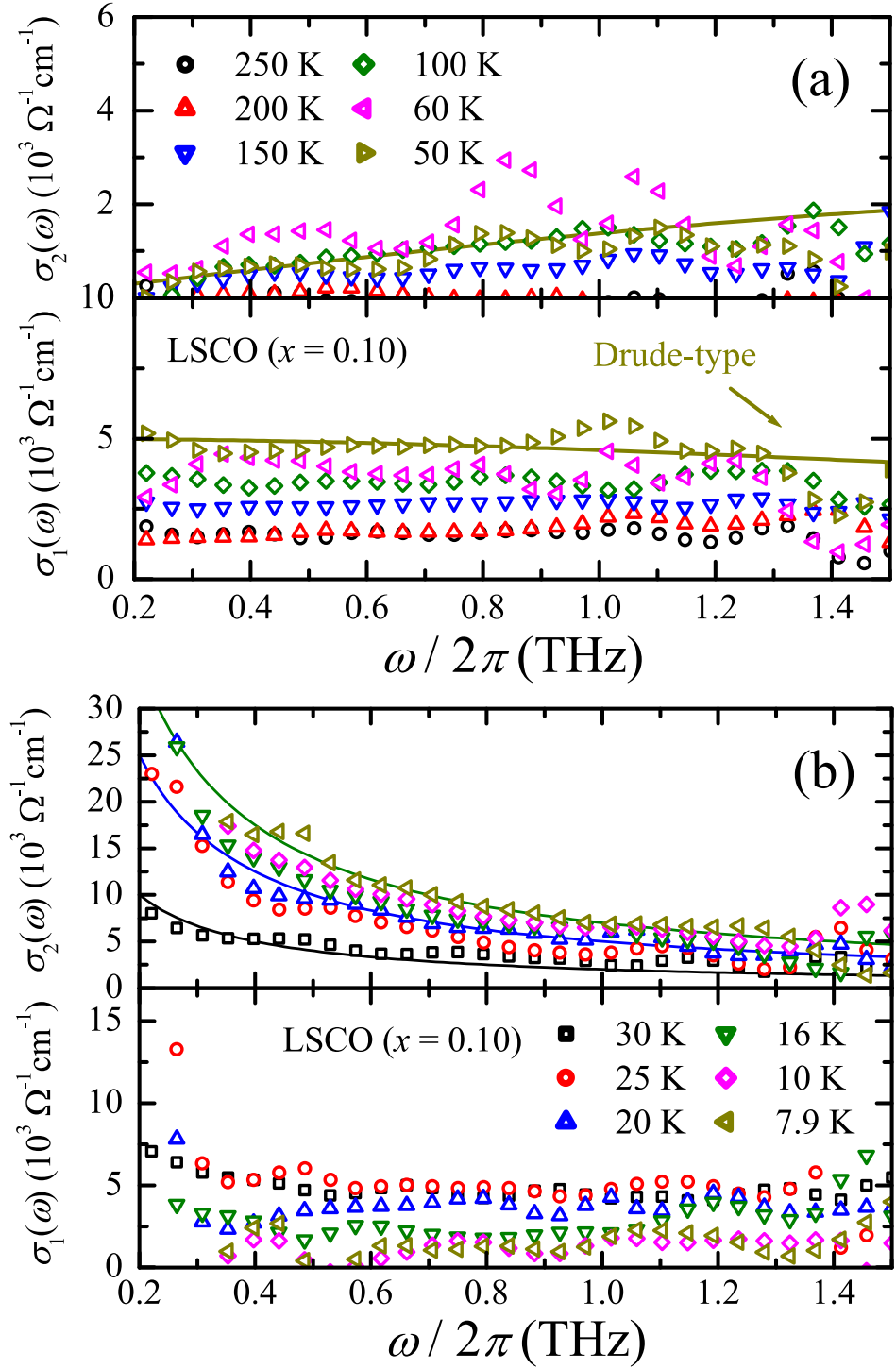}
\caption{\label{fig:x010sigma} (Color online) Frequency dependence of the complex conductivity of the underdoped LSCO ($x$ = 0.10) film (a) in the normal state ($T \geq  $ 50 K) and (b) near the superconducting state ($T \leq $ 30 K).
The lower panels are the real part, and the upper panels are the imaginary part of $\tilde \sigma (\omega)$, respectively.
In (a), the solid lines are the fitting to the Drude conductivity ($\tau$ = 0.05 ps) of the data at $T$ = 50 K.
In (b), the solid lines in the $\sigma_2$ panel are the fitting curves ($\sigma_2 \propto \omega^{-1}$) against the data at $T$ = 30 K, 20 K, and 7.9 K.}
\end{center}
\end{figure}

\begin{figure*}[!t]
\begin{center}
\includegraphics[width=0.99\linewidth]{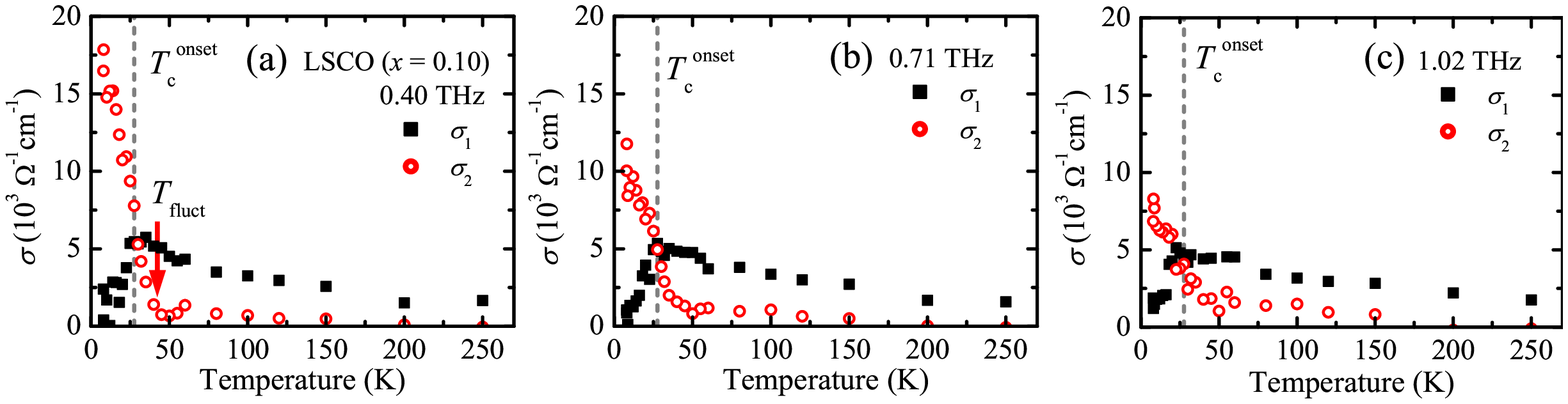}
\caption{\label{fig:x010sigmaW} (Color online) Temperature dependence of the complex conductivity ($\tilde \sigma (T)$) of the underdoped LSCO ($x$ = 0.10) film at (a) 0.40 THz, (b) 0.71 THz, and (c) 1.02 THz. In these figures, $\sigma_1(T)$ are closed squares and $\sigma_2(T)$ are open circles. The broken lines are $T_\text{c}^\text{onset}$. The down arrow in (a) indicates the onset temperature of superconducting fluctuation, $T_\text{fluct}$.}
\end{center}
\end{figure*}

In Fig. \ref{fig:x010sigma}, we show the complex conductivity spectrum, $\tilde \sigma (\omega)$, of the LSCO ($x$ = 0.10) film ((a) in the normal state ($T \geq  $ 50 K) and (b) near the superconducting state ($T \leq $ 30 K)).
In Figs. \ref{fig:x010sigma}(a) and \ref{fig:x010sigma}(b), the lower panel is the real part, $\sigma_1$, and the upper panel is the imaginary part, $\sigma_2$.
$\sigma_1(\omega \to 0)$ is approximately $1.6 \times 10^3 \ \Omega^{-1} \text{cm}^{-1}$ at $T$ = 250 K, which is close to the dc conductivity of LSCO single crystals ($\sigma_\text{dc} \sim 1.5 \times 10^3 \ \Omega^{-1} \text{cm}^{-1}$ at $T$ = 250 K for $x$ = 0.10).\cite{Ando2004PRLv2}
Therefore, we believe that the LSCO thin film used in this work has a similar quality to good LSCO single crystals.
In the normal state, $\tilde \sigma (\omega)$ is that of a good metal; $\sigma_1$ is almost frequency independent, and $\sigma_2$ is proportional to $\omega$.
Indeed, the solid lines in Fig. \ref{fig:x010sigma}(a) are the fitting to the Drude model ($\tau$ = 0.05 ps) of the data at $T$ = 50 K.
$\tilde \sigma (\omega)$ is well fit by this Drude-type spectrum.
With decreasing temperature, the frequency-dependent feature is more pronounced even in $\sigma_1$, which is due to the enhancement of $\tau$ at low temperatures.
In contrast, in the superconducting state (Fig. \ref{fig:x010sigma}(b)), $\sigma_1(\omega)$ is suppressed compared with $\sigma_1(\omega)$ at $T$ = 25-30 K in the entire frequency region, due to the gap opening in the excitation spectrum, and $\sigma_2(\omega)$ is proportional to the inverse of frequency, which is the result of the delta function in $\sigma_1$ at zero frequency.
The solid lines in the $\sigma_2$ panel are the fitting curves ($\sigma_2 \propto \omega^{-1}$) against the data at $T$ = 30 K, 20 K, and 7.9 K.
The magnetic penetration depth, $\lambda$, can be calculated from $\sigma_2$ as $\lambda = (\sigma_2 \mu_0 \omega)^{-1/2}$, where $\mu_0$ is the magnetic permeability of a vacuum.
Assuming the two-fluid model ($\lambda (T) = \lambda (0) \sqrt{1-(T/T_\text{c})^4}$), we can estimate that $\lambda (0) = 0.44 \ \mu$m in the $x$ = 0.10 sample, which is in good agreement with a previous microwave result.\cite{Shibauchi1994PRL}

Figure \ref{fig:x010sigmaW} is the temperature dependence of the complex conductivity, $\tilde \sigma (T)$, of the LSCO ($x$ = 0.10) film at (a) 0.40 THz, (b) 0.71 THz, and (c) 1.02 THz.
$\sigma_1(T)$ are the closed squares and $\sigma_2(T)$ are the open circles.
$\sigma_1(T)$ monotonically decreases from a temperature around $T_\text{c}^\text{onset}$ (27.6 K, gray broken lines).
The value of $\sigma_1(T)$ extrapolated to $T$ = 0 K is $1.0 \times 10^3 \ \Omega^{-1}\text{cm}^{-1}$, which suggests that most of the normal carrier condensed into superfluid at absolute zero.
We should note that there is almost no frequency dependence in $\sigma_1 (T)$ at 0.40 THz, 0.71 THz, and 1.02 THz in the superconducting state.
We will discuss this behavior later.
Conversely, $\sigma_2 (T)$ has an obvious frequency dependence in the entire temperature range.
$\sigma_2 (T)$ starts to rapidly increase from 40-45 K, which can be clearly seen in Fig. \ref{fig:x010sigmaW}(a).
Therefore, we determined the onset temperature of superconducting fluctuation, $T_\text{fluct}$, as 42.5 K (the down arrow) for this film.
One should note that the finite $\sigma_2(T)$ value above 50 K does not originate from superconducting fluctuation, because the frequency spectrum is the Drude-type, as shown in Fig. \ref{fig:x010sigma}(a).
A previous THz conductivity study of Bi cuprate\cite{Murakami2002JS} insisted that superconducting fluctuation starts from $\sim$ 210 K because of the finite $\sigma_2$.
We believe this to be the contribution of the Drude-type normal carrier.

\begin{figure}[!t]
\begin{center}
\includegraphics[width=0.99\linewidth]{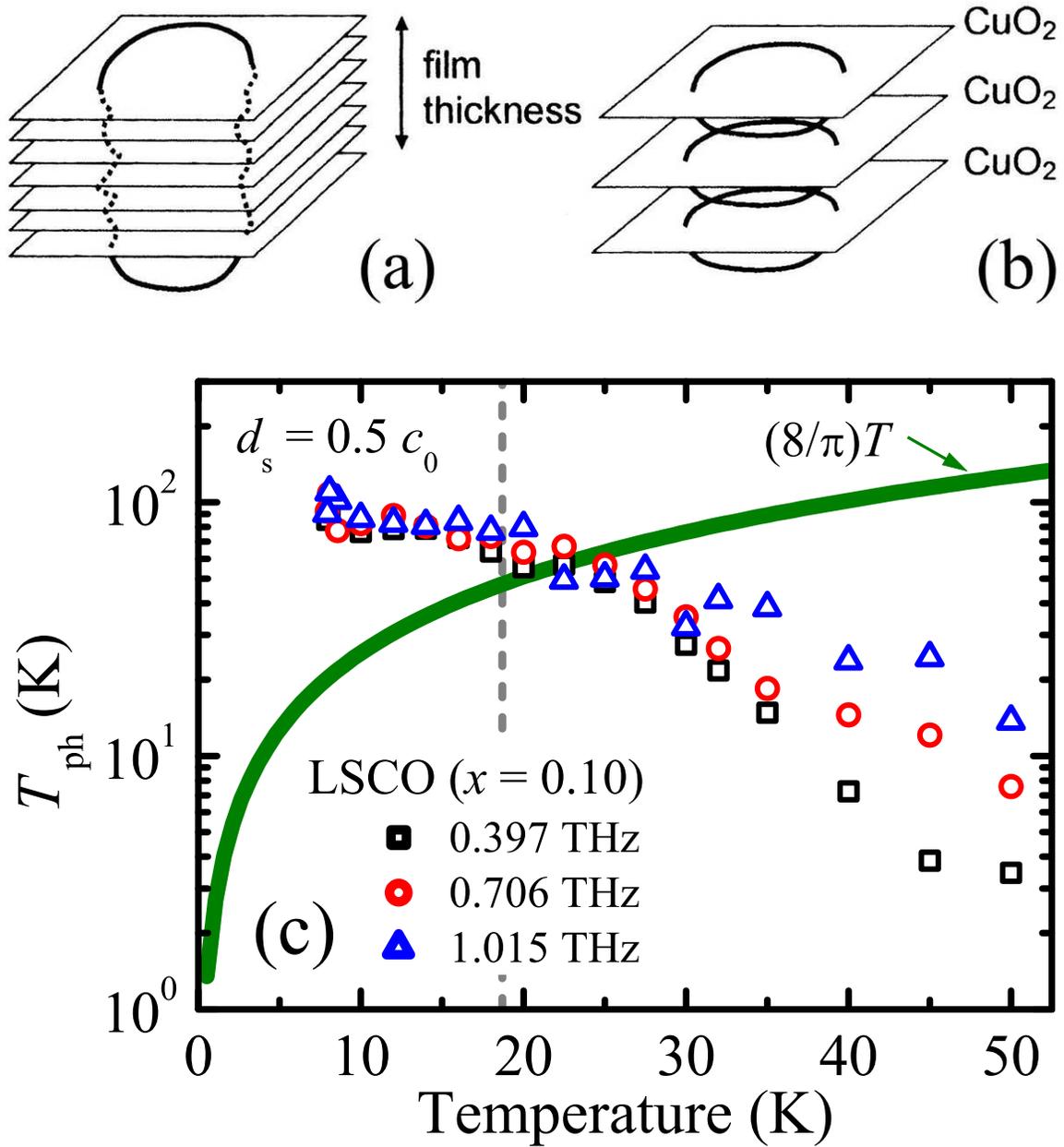}
\caption{\label{fig:x010Ttheta} (Color) (a)(b) Schematic figures of the vortex-antivortex pair, (a) when $d_\text{s}$ is larger than the interdistance of the CuO$_2$ plane and is almost equal to the film thickness, and (b) when $d_\text{s}$ is comparable to the interdistance of the CuO$_2$ plane.
(c) Temperature dependence of the characteristic temperature, $T_\text{ph} (T)$, of the LSCO ($x$ = 0.10) film at several frequencies. $d_\text{s}$ is set as 0.5$c_0$. The broken line (gray) is $T_\text{c}^\text{zero}$, and the solid line (green) represents $(8/\pi )T$, which is the value of $T_\theta$ when the universal jump takes place at $\omega = 0$ in the case of the BKT transition. Details are described in the text.}
\end{center}
\end{figure}

Next, we discuss the temperature dependence of the phase stiffness temperature, $T_\theta (T)$.
Because extracting only the superfluid contribution from experimentally obtained $\sigma_2$ is sensitive to the method of analysis, we calculated a characteristic temperature, $T_\text{ph}$, as $\frac{\hbar}{k_B e^2} \hbar \omega \sigma_2 d_s$, from the imaginary part of the measured complex conductivity.
At high temperatures above $T_\text{c}$, $T_\text{ph} $ exceeds $T_\theta$ in general because the contribution of Drude conductivity to $\sigma_2$ becomes non-negligible.
On the other hand, $T_\text{ph}$ becomes equivalent to $T_\theta$, when $\sigma_2$ comes from superfluid alone at low temperatures.
The microwave conductivity measurement\cite{Kitano2006PRB} previously confirmed that underdoped LSCO shows a BKT-type $T_\theta (T)$ and that $T_\text{BKT}$ was close to $T_\text{c}^\text{zero}$.
In the definition of  $T_\theta$ (Eq. (\ref{eq:BKT})), $d_\text{s}$ represents the size of the vortex-antivortex pair in superconductor.
One can consider $d_s$ in two different ways, which are represented in Figs. \ref{fig:x010Ttheta}(a) and \ref{fig:x010Ttheta}(b) schematically.
One way is that $d_\text{s}$ is far larger than the interdistance of the CuO$_2$ plane and can even take about the film thickness (Fig. \ref{fig:x010Ttheta}(a)).
The other is that $d_\text{s}$ is comparable to the interdistance of the CuO$_2$ plane (Fig. \ref{fig:x010Ttheta}(b)).
For materials with a BKT-like superconducting fluctuation, we can estimate explicitly which consideration is valid by comparing the $T_\text{ph} (T_\text{BKT})$ data with $(8/\pi )T_\text{BKT}$.

Figure \ref{fig:x010Ttheta}(c) shows $T_\text{ph} (T)$ at 0.40 THz, 0.71 THz, and 1.02 THz.
The broken line (gray) in Fig. \ref{fig:x010Ttheta}(c) is $T_\text{c}^\text{zero}$, and the solid line (green) shows $(8 / \pi) T$, which is the value of $T_\theta$ when the universal jump takes place at $\omega$ = 0 by the BKT transition.
We cannot observe the obvious universal jump of $T_\text{ph}$ at $T_\text{c}^\text{zero}$ because of the high frequency, which is the same tendency observed in BSCCO\cite{Corson1999nature} and even in the microwave region of LSCO.\cite{Kitano2006PRB}
Although a clear jump is absent, we can still estimate $d_s$ from $T_\text{ph} (T_\text{c}^\text{zero})$ because it is found to be the same magnitude as $(8 / \pi) T_\text{c}^\text{zero}$ (the intersection point between the gray broken line and the green solid line) if we set $d_\text{s} = 0.5 c_0$ ($c_0$ = 13.216 \AA $\ $for the LSCO ($x$ = 0.10) film).
The factor 0.5 corresponds to the fact that two CuO$_2$ planes exist in the LSCO unit cell.
This suggests that the vortex-antivortex pair makes a loop in a single CuO$_2$ plane (Fig. \ref{fig:x010Ttheta}(b)).
Though there is a slight deviation between $T_\text{ph} (T)$ and $(8/\pi )T$ at $T = T_\text{c}^\text{zero}$ due to the error in the estimation of $d_\text{s}$ or that of $\sigma_2$, our THz data ($T_\text{ph} (\omega, T)$) resembles to that of the microwave study,\cite{Kitano2006PRB} which strongly suggests that the BKT phase transition takes place in the underdoped LSCO film.
We therefore conclude that the $T_\text{ph}$ behavior found in underdoped LSCO ($x$ = 0.10) film originates from the two-dimensional superconducting fluctuation nature.

\begin{figure*}[!t]
\begin{center}
\includegraphics[width=0.99\linewidth]{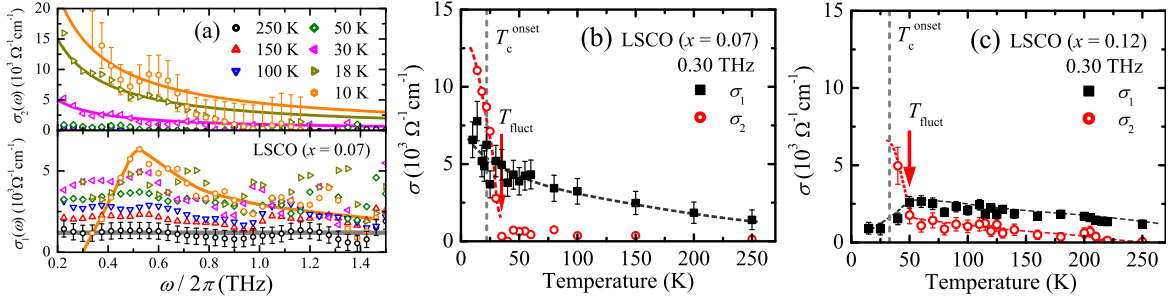}
\caption{\label{fig:x007sigmaW} (Color online) (a) $\tilde \sigma (\omega)$ of the underdoped LSCO ($x$ = 0.07) film. The lower panel is the real part ($\sigma_1$), and the upper panel is the imaginary part ($\sigma_2$). The solid line in the $\sigma_1$ panel indicates dc conductivity (1.2 $\times 10^3\ \Omega^{-1} $cm$^{-1}$) at $T$ = 250 K and the guideline for the data at 10 K. The solid line in the $\sigma_2$ panel indicates $\omega^{-1}$ fitting curves for the data at $T$ = 30 K, 18 K, and 10 K.
(b)(c) Temperature dependence of $\sigma_1$(closed squares) and $\sigma_2$(open circles) at 0.30 THz of the underdoped LSCO ((b) $x$ = 0.07, (c) $x$ = 0.12) films. In (b) and (c), the broken lines are $T_\text{c}^\text{onset}$ (gray) and the guide to the eye for $\sigma_1 (T)$ (black) and $\sigma_2 (T)$ (red). The down arrows indicate the $T_\text{fluct}$ of each film.}
\end{center}
\end{figure*}

Next, we show the results of another underdoped LSCO ($x$ = 0.07) film ($T_\text{c}^\text{onset}$ = 22.1 K, $T_\text{c}^{\text{zero}}$ = 18.5 K).
Figure \ref{fig:x007sigmaW}(a) shows the $\tilde \sigma (\omega)$ of this film.
In the $\sigma_1 (\omega)$ panel, the gray solid line is $\sigma_\text{dc}$ (1.2 $\times 10^3\ \Omega^{-1} $cm$^{-1}$) at 250 K, and the orange solid line is the guide to the eye for the data at 10 K.
In the normal state, a Drude-type spectrum ($\sigma_1 \sim \sigma_\text{dc}, \sigma_2 \sim 0$) was observed, which is similar to the LSCO ($x$ = 0.10) film.
For example, $\sigma_1 (\omega)$ is almost independent of frequency and corresponds to $\sigma_{\text{dc}}$ within the error-bar at $T$ = 250 K.

In the superconducting state, we obtained $\lambda(0) = 0.5 \ \mu$m from the slope of the $\omega^{-1}$ fitting curves.
In addition, we observed a hump structure in $\sigma_1(\omega)$ ($\sim$ 0.5 THz at 10 K) not observed in the LSCO ($x$ = 0.10) film.
Possible origins of this hump structure may be (1) the superconducting gap or (2) the Josephson plasma because both phenomena have similar energy scales and occur only in the superconducting state.
In case of (1), we can regard 2$\Delta$ as 2.05 meV (0.5 THz) at $T$ = 10 K, where $2\Delta$ is the superconducting gap energy.
Conversely, if we set $T_\text{c}$ as 20 K ($\sim T_\text{c}^\text{zero}$), the superconducting gap energy at absolute zero, $\Delta (0)$, is calculated to be 1.09 meV, using the empirical temperature dependence of the BCS-type superconductor for the superconducting gap energy, $\Delta (T)  =\Delta (0) \tanh (1.74 \sqrt{T_\text{c}/T -1})$.
This leads to $2\Delta (0) / k_\text{B} T_\text{c}$ = 1.2, which is much smaller than the BCS value of 3.53.
Therefore, we believe that a hump structure in $\sigma_1(\omega)$ does not originate from superconducting gap energy.
For (2), the Josephson plasma resonance is widely known to be observed in the infrared conductivity along the $c$-axis, which causes a similar hump structure in the conductivity spectrum.\cite{Tamasaku1992PRL}
The $c$-axis THz reflectivity measurement of underdoped LSCO ($x$ = 0.10) single crystal\cite{Matsuoka2009PC} shows a characteristic reflectivity structure at 22.5 cm$^{-1}$ ($\sim$ 0.7 THz), which arises from Josephson plasma resonance.
Because 0.7 THz is close to the frequency-scale of a hump structure in $\sigma_1(\omega)$ for our LSCO ($x$ = 0.07) film, we assume that the $c$-axis component contributes to a hump structure in $\sigma_1(\omega)$ caused by the slight deviation of alignment between the direction of the electric field of the THz pulse and the $ab$ plane of LSCO, in the LSCO ($x$ = 0.07) film experiment.
For example, some groups have reported that a strong anisotropy of HTSC result in an anomalous conductivity spectrum.\cite{Pimenov2002PRB, Tajima2005PRB}
The strong anisotropic nature of HTSC also sensitively affects $\sigma_1$ structure in this film.
We stress that the onset temperature of the superconducting fluctuation, $T_\text{fluct}$, is not changed by the additional small $c$-axis component.

Other underdoped LSCO films also show qualitatively similar temperature dependence of $\sigma_2$.
For LSCO ($x$ = 0.07) film, we show $\tilde \sigma (T)$ at 0.30 THz in Fig. \ref{fig:x007sigmaW}(b).
The broken lines are $T_\text{c}^\text{onset}$ (gray) and the guidelines for $\sigma_1 (T)$ (black) and $\sigma_2 (T)$ (red).
$\sigma_2 (T)$ (open circles) qualitatively shows similar results to those of the LSCO ($x$ = 0.10) film, in the sense that $\sigma_2 (T)$ rapidly increases far above $T_\text{c}^\text{zero}$.
$T_\text{fluct}$ can be estimated as about 35 K (the down arrow), which corresponds to $T_\text{fluct}/T_\text{c}^\text{zero} \sim 2$.
This result is consistent with the conclusion obtained in a previous analysis of microwave conductivity data.\cite{Ohashi2009PRB}
$\sigma_1 (T)$ (closed squares) does not clearly decrease to zero at $T \to 0$, which seems to be influenced by the hump structure in $\sigma_1 (\omega)$, though detailed reason for this remains undetermined.

\begin{figure*}[!t]
\begin{center}
\includegraphics[width=0.99\linewidth]{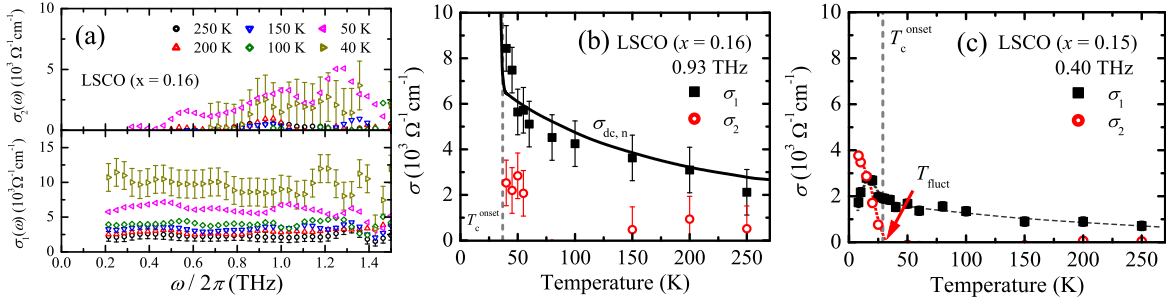}
\caption{\label{fig:x016WF} (Color online) (a) $\tilde \sigma (\omega)$ in the normal state of the optimally doped LSCO ($x$ = 0.16) thin film. The lower panel is $\sigma_1 (\omega)$, and the upper panel is $\sigma_2 (\omega)$. (b) $\tilde \sigma (T)$ of LSCO ($x$ = 0.16) film at 0.93 THz. The closed squares are $\sigma_1 (T)$, and the open circles are $\sigma_2(T)$. The black solid line is the normalized dc conductivity, $\sigma_\text{dc,n} = 1.5\sigma_\text{dc}$, and the gray broken line is $T_\text{c}^\text{onset}$. (c) $\tilde \sigma (T)$ of LSCO ($x$ = 0.15) film at 0.40 THz. The symbols are same as (b), and the down arrow shows $T_\text{fluct}$ (30 K).}
\end{center}
\end{figure*}

For LSCO ($x$ = 0.12) film ($T_\text{c}^\text{onset} = 32.9$ K, $T_\text{c}^\text{zero} = 27.9$ K), $\tilde \sigma (T)$ at 0.30 THz is depicted in Fig. \ref{fig:x007sigmaW}(c).
There, $\sigma_1 (T)$ (closed squares) increases with decreasing temperature in the normal state and approaches zero in the superconducting state.
$\sigma_2 (T)$ (open circles) starts to rapidly increase from about 50 K ($T_\text{fluct}$, the down arrow).
We observed a rather larger value of  $\sigma_2$ compared to those of the other underdoped LSCO ($x$ = 0.10, 0.07) films in the normal state.
This is probably due to the increase in carrier lifetime ($\tau$) with increasing carrier concentration, which was also observed in the optical conductivity.\cite{Suzuki1989PRB}

\subsection{\label{sec:level3-2}optimally doped region: 0.15 $\leq x \leq$ 0.16}

For the optimally doped region, we first show the results of the LSCO ($x$ = 0.16) film ($T_\text{c}^\text{onset}$ = 36.9 K, $T_\text{c}^\text{zero}$ = 32.0 K).
Figure \ref{fig:x016WF}(a) is $\tilde \sigma (\omega)$ in the normal state (lower panel: $\sigma_1 (\omega)$, upper panel: $\sigma_2 (\omega)$).
$\tilde \sigma (\omega)$ represents a Drude-like spectrum at $T \geq $ 40 K, and we cannot observe an increase in $\sigma_2$ at low frequencies.
Therefore, we can conclude that the superconducting fluctuation starts at least below 40 K in the LSCO ($x$ = 0.16) film.
Figure \ref{fig:x016WF}(b) is $\tilde \sigma (T)$ of this film at 0.93 THz.
The closed squares are $\sigma_1 (T)$, and the open circles are $\sigma_2(T)$.
The black solid line is the normalized dc conductivity, $\sigma_\text{dc,n} = 1.5\sigma_\text{dc}$, and the gray broken line is $T_\text{c}^\text{onset}$.
$\sigma_1(T)$ in the normal state is close to $\sigma_\text{dc,n}$, suggesting a slight difference between $\sigma_1 (T)$ and $\sigma_\text{dc} (T)$.
This difference may come from an error in the estimation of film thickness, the slight $c$-axis component in the dc measurement, or other sources.

We previously reported the results of another optimally doped LSCO ($x$ = 0.15) film ($T_\text{c}^\text{onset}$ = 29.0 K, $T_\text{c}^\text{zero}$ = 25.0 K).\cite{Nakamura2009M2S,Maeda2010PhysC}
In those studies, we found that $\sigma_2$ starts to increase from $T/ T_\text{c}^\text{zero} = 1.2$.
In Fig. \ref{fig:x016WF}(c), we show $\tilde \sigma (T)$ at 0.40 THz.
We observed a broad peak structure in $\sigma_1 (T)$ (closed squares) in the superconducting state, and $\sigma_1 (T)$ then decreases at $T \to 0$.
$\sigma_2 (T)$ (open circles) starts to increase from approximately 30 K ($T/ T_\text{c}^\text{zero} = 1.2$), which is shown by the down arrow at $T_\text{fluct}$.
This behavior can be more clearly seen in the upper panel of Fig. \ref{fig:sigmaT_all}(b), which will be discussed later.

\begin{figure*}[!t]
\begin{center}
\includegraphics[width=0.99\linewidth]{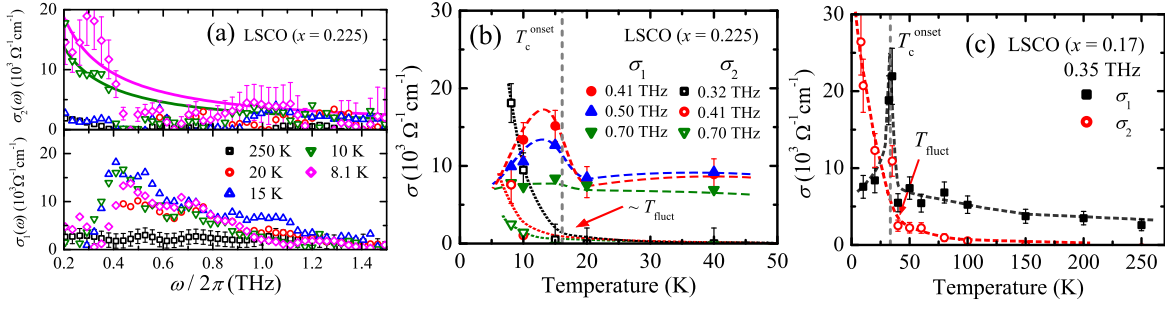}
\caption{\label{fig:x0225sigma} (Color online) (a) $\tilde \sigma (\omega)$ of the overdoped LSCO ($x$ = 0.225) film.
The solid lines in the $\sigma_2$ panel are the $\omega^{-1}$ guidelines for the data at 10 K and 8.1 K.
(b) $\tilde \sigma (T)$ of the overdoped LSCO ($x$ = 0.225) film at several frequencies. (c) $\tilde \sigma (T)$ of the overdoped LSCO ($x$ = 0.17) film at 0.35 THz. In (b) and (c), the closed symbols are $\sigma_1$, the open symbols are $\sigma_2$, the gray broken lines are $T_\text{c}^\text{onset}$, and other broken lines are guides for the eye. The arrows indicate $T_\text{fluct}$ of each film.}
\end{center}
\end{figure*}

\subsection{\label{sec:level3-2}overdoped region: $x \geq 0.17$}

Figure \ref{fig:x0225sigma}(a) is $\tilde \sigma (\omega)$ of the LSCO ($x$ = 0.225) film ($T_\text{c}^\text{onset}$ = 16.1 K, $T_\text{c}^\text{zero}$ = 10.9 K).
$\sigma_1$ is the lower panel, and $\sigma_2$ is the upper panel.
In the $\sigma_2$ panel, the solid lines are the $\omega^{-1}$ guidelines for the data at 10 K and 8.1 K.
We observed a hump structure in $\sigma_1 (\omega)$ around 0.4 THz, similar to the underdoped LSCO ($x$ = 0.07) film.
This structure does not originate from the superconducting gap, because $2\Delta(0) / k_\text{B} T_\text{c}^\text{zero} = 2.0$ is far smaller than the BCS value if we set $2\Delta$ (8.1 K) = 0.35 THz (1.4 meV).
Therefore, similar to the LSCO ($x$ = 0.07) film, we suppose that the Josephson plasma resonance of a very small mixture of $c$-axis conductivity causes a hump structure in $\sigma_1 (\omega)$ in this film.
Fig. \ref{fig:x0225sigma}(b) shows $\tilde \sigma (T)$ with several frequencies, where $\sigma_1$ is the closed symbol, and $\sigma_2$ is the open symbol, the gray broken line is $T_\text{c}^\text{onset}$, and the other broken and dotted lines indicate the guidelines for $\sigma_1$ and $\sigma_2$, respectively..
We observed a qualitatively different $\sigma_1$ behavior from the underdoped and the optimally doped films.
For the low-frequency region (the data at 0.41 THz (red) and 0.50 THz (blue)), the relatively sharp peak structure is observed in $\sigma_1 (T)$ near $T_\text{c}$.
In Fig. 10(c), we can see this behavior more clearly.
In contrast, in the high-frequency region (the data at 0.70 THz (green)), the peak structure is obscure and $\sigma_1 (T)$ almost does not change in the superconducting state.
In other words, there is some residual conductivity.
We will discuss this point later.
From $\sigma_2 (T)$, we roughly estimated $T_\text{fluct}$ as $\sim$15-20 K in this LSCO film (the arrow in Fig. \ref{fig:x0225sigma}(b)).

\begin{figure*}[!t]
\begin{center}
\includegraphics[width=0.99\linewidth]{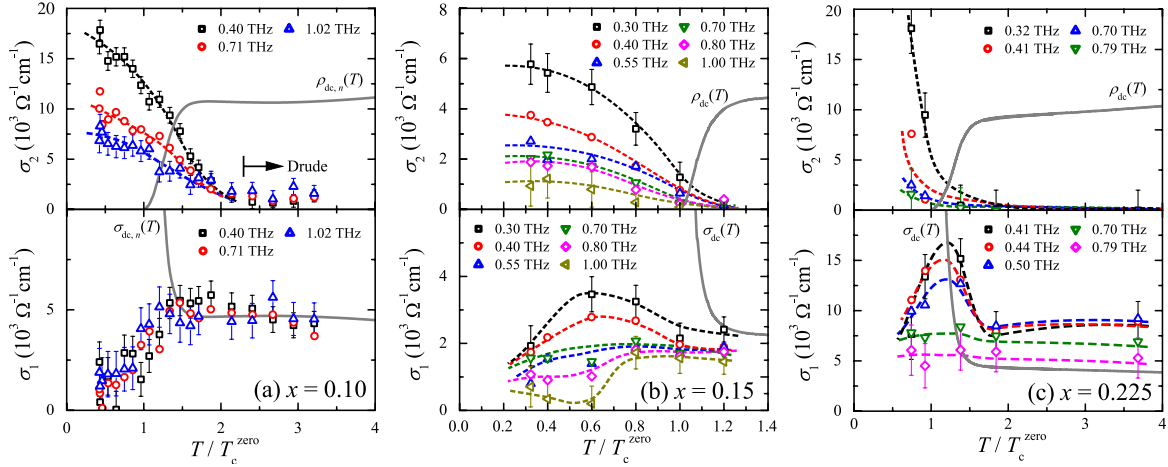}
\caption{\label{fig:sigmaT_all} (Color online). 
Temperature dependence of the complex conductivity of (a) the underdoped ($x$ = 0.10), (b) the optimally doped ($x$ = 0.15), and (c) the overdoped ($x$ = 0.225) LSCO films near $T_\text{c}$. The lower panels are the real part, $\sigma_1$, and the upper panels are the imaginary part, $\sigma_2$. The horizontal axis is the reduced temperature by $T_\text{c}^\text{zero}$. 
The broken lines are guides for the eye. The solid lines in the $\sigma_1$ and $\sigma_2$ panels are the dc conductivity and the dc resistivity, respectively.
\vspace{-0.5cm}}
\end{center}
\end{figure*}

Another overdoped LSCO ($x$ = 0.17) film ($T_\text{c}^\text{onset}$ = 33.4 K, $T_\text{c}^\text{zero}$ = 30.3 K) shows similar results to the LSCO ($x$ = 0.225) film.
In $\tilde \sigma (T)$ (0.35 THz, Fig. \ref{fig:x0225sigma}(c)), we found a sharp peak structure in $\sigma_1 (T)$ at $T_\text{c}$ that is similar to the LSCO ($x$ = 0.225) film.
Therefore, we suggest that the $\sigma_1(T)$ peak structure at $T_\text{c}$ is a common aspect for overdoped LSCO.
From $\sigma_2 (T)$, we can estimate $T_\text{fluct}$ as $\sim$ 40 K in this sample (the arrow in Fig. \ref{fig:x0225sigma}(c)).

Regarding the two characteristic aspects of $\tilde \sigma (T)$ mentioned above ($\sigma_1 (T)$ peak near $T_\text{c}$ and residual conductivity), we first comment on the residual conductivity.
The origin of the residual conductivity is suggested to be (1) the result of chemical instability during the synthesis process of LSCO for the large Sr concentration or (2) an intrinsic microscopic phase separation between the superconducting state (paired carriers) and the normal state (unpaired carriers).\cite{Uemura1993Nature}
With the magnetization of LSCO single crystals, the volume fraction of the superconducting phase gradually decreases with increasing Sr concentration in the overdoped region,\cite{Tanabe2005JPSJ, Adachi2006PC} which supports the hypothesis (2).
In addition, as we described in the introduction, the microwave conductivity analysis of LSCO\cite{Ohashi2009PRB} indicated that the critical dynamics of the superconducting fluctuation is two-dimensional and has different critical indices, which are often observed for a number of frustrated $XY$-models.\cite{Hasenbach2005JSM}
This may also support hypothesis (2).
In addition, in the overdoped YBCO, a far-infrared $c$-axis conductivity study reported that the residual conductivity had a relatively large value compared with other doping levels, which also suggests the coexistence of paired and unpaired carriers in the superconducting state of overdoped HTSC.\cite{Schutzmann1994PRL}
Therefore, we considered that the residual conductivity in our overdoped LSCO film resulted from an intrinsic microscopic heterogeneity, characteristic of overdoped HTSC.

Next, we discuss the origin of the $\sigma_1(T)$ peak near $T_\text{c}$.
Similar sharp peak structure of $\sigma_1(T)$ near $T_\text{c}$ was also observed in the microwave conductivity of YBCO.\cite{Olsson1991PhysC, Anlage1996PRB, Hosseini1999PRB}
There, the $\sigma_1$ peak can be fitted by the two-fluid model\cite{Olsson1991PhysC} or the effective medium model,\cite{Anlage1996PRB} which assume some spatial distribution of $T_\text{c}$ in the sample.
However, with very high-quality samples, Hosseini $et \ al.$ demonstrated that the $\sigma_1$ peak of YBCO near $T_\text{c}$ comes from superconducting fluctuation rather than the spatial distribution of $T_\text{c}$.\cite{Hosseini1999PRB}
Therefore, we suggest that the $\sigma_1(T)$ peak near $T_\text{c}$, obtained in our overdoped LSCO film also comes from some characteristics in the overdoped HTSC, superconducting fluctuation and an intrinsic microscopic phase separation, for example.
For the detailed analysis for this issue, the microspectroscopic method will be needed .

\subsection{Doping dependence}

We consider all the results together and make some general remarks regarding the doping dependence.
Figure \ref{fig:sigmaT_all} is $\tilde \sigma (T)$ of (a) the underdoped ($x$ = 0.10), (b) the optimally doped ($x$ = 0.15), and (c) the overdoped ($x$ = 0.225) LSCO films near $T_\text{c}$.
The lower panels are $\sigma_1$, and the upper panels are $\sigma_2$.
The horizontal axis is the reduced temperature by $T_\text{c}^\text{zero}$ of each film, the broken lines are the guides for the eye, and the gray solid lines in the $\sigma_1$ and $\sigma_2$ panels are the dc conductivity and the dc resistivity, respectively (for $x$ = 0.10, the dc value is normalized).

For $\sigma_1 (T)$, we found qualitatively different features concerning the following three aspects with changing doping levels: (1) the broad peak structure at $T\sim T_\text{c}/ 2$, (2) the relatively sharp peak near $T_\text{c}$, and (3) the residual conductivity at $T \to 0$.
In the underdoped LSCO film (the lower panel of Fig. \ref{fig:sigmaT_all}(a)), $\sigma_1 (T)$s with different frequencies are almost the same, even in the superconducting state.
Only in the optimally doped sample (the lower panel of Fig. \ref{fig:sigmaT_all}(b)) did we observe a broad peak structure in the superconducting state ($T \sim T_\text{c}/2$).
The broad peak structure in $\sigma_1 (T)$ clearly depends on frequency.
At low frequencies (e.g., 0.30 THz), $\sigma_1 (T)$ gradually increases with decreasing temperature from approximately $T_\text{c}^\text{onset}$, and there is a broad peak near $T / T_\text{c}^\text{zero} \sim 0.6$.
On the other hand, at high frequencies (e.g., 1.00 THz), the magnitude of the broad peak in $\sigma_1(T)$ is reduced, and the peak temperature at which $\sigma_1(T)$ takes a maximum value approaches $T_\text{c}$.
We suggest that the $\sigma_1$ broad peak structure found only in the optimally doped LSCO film relates to the dimensionality of the critical dynamics of HTSC.
As we described in the introduction, a dynamical scaling analysis of microwave conductivity in LSCO showed that the critical dynamics becomes three-dimensional only in the optimally doped region,\cite{Ohashi2009PRB}.
In addition, we could not find $\sigma_1$ broad peak structure in our THz data (Fig. \ref{fig:x010sigmaW}) or in the microwave conductivity data\cite{Kitano2006PRB} of the underdoped LSCO, whose critical dynamics is a two-dimensional $XY$ type (BKT transition).
This indicates to relate the $\sigma_1$ broad peak structure with three-dimensional critical dynamics.
Indeed, there are similar and consistent experimental results even in other HTSCs.
For YBCO, a broad $\sigma_1 (T)$ peak is found in the microwave region,\cite{Hosseini1999PRB} whose critical dynamics is the three-dimensional $XY$ type.\cite{Kamal1994PRL}
Such a broad $\sigma_1 (T)$ peak is also found in THz region,\cite{Nuss1991PRL} which is thought to result from both the decrease in concentration of the unpaired carrier and the increase in carrier lifetime with decreasing temperature.
For BSCCO, the broad $\sigma_1 (T)$ peak at $T\sim T_\text{c} / 2$ is not found in the THz conductivity measurement of underdoped BSCCO,\cite{Corson1999nature} whose critical dynamics is a two-dimensional.
Therefore, we suggest that the broad $\sigma_1(T)$ peak at $T\sim T_\text{c} / 2$ is characteristic of the superconductor, whose critical dynamics is three-dimensional.
In the overdoped LSCO ($x$ = 0.225) film (lower panel of Fig. \ref{fig:sigmaT_all}(c)), we observed a relatively sharp $\sigma_1 (T)$ peak structure at near $T_\text{c}$, and some residual conductivity in the superconducting state.
As discussed in Sec. 3.3, these types of characteristic behavior are thought to arise from superconducting fluctuation and the intrinsic microscopic phase separation of overdoped HTSC.
We note that the strong doping dependence of the transient relaxation time of the anti-nodal quasiparticle, investigated by the pump-prove method in BSCCO\cite{Gedik2005PRL} is also related to these dimensional changes in critical dynamics, obtained in this work.
There, the relaxation time for the optimally doped BSCCO sample was smaller than other doping levels, which suggests that the fluctuation effect is small for the optimally doped region.
In addition, the photoinduced change in reflectivity by the pump pulse is opposite in sign between the underdoped and overdoped BSCCO, which also suggests that the nature of the quasiparticle dynamics is very sensitive to doping level.

\begin{figure}[!t]
\begin{center}
\begin{minipage}{0.9\linewidth}
\includegraphics[width=0.99\linewidth]{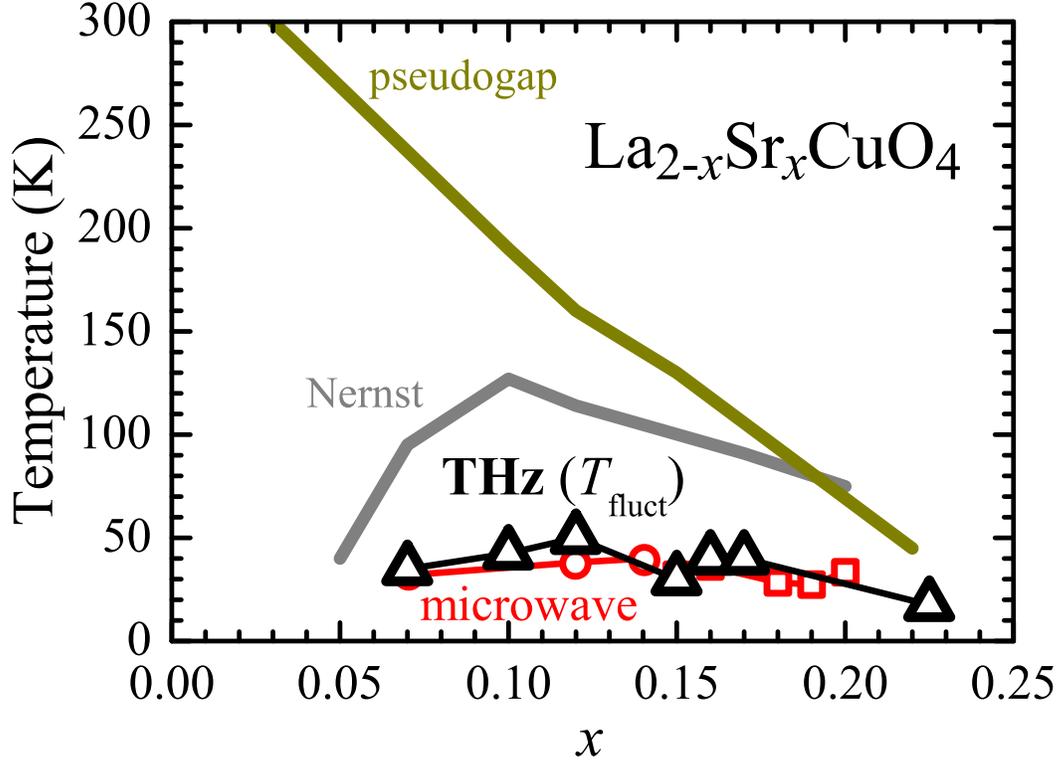}
\end{minipage}
\caption{\label{fig:PD} (Color) The electronic phase diagram of LSCO.
The open triangles show the onset temperature of the superconducting fluctuation obtained in this work, $T_\text{fluct}$, which is defined by the starting temperature of the rapid increase of $\sigma_2 (T)$ by the superfluid.
The open circles and squares show the corresponding results of the microwave study,\cite{Ohashi2009PRB} where the analytically extrapolated values were used for the underdoped region (circles).
The solid lines are the onset temperature of the Nernst signal (gray) \cite{Wang2006PRB} and pseudogap temperature observed by photoemission spectroscopy (olive).\cite{Hashimoto2007PRB}}
\end{center}
\end{figure}

Next, for $\sigma_2 (T)$, there is a qualitatively common aspect among different doping levels.
$\sigma_2 (T)$ starts to increase rapidly at a temperature far above $T_\text{c}^\text{zero}$ due to superconducting fluctuation.
However, we stress that the starting temperature of the superconducting fluctuation, $T_\text{fluct}$, is quantitatively different among LSCO films with different doping levels, in the sense of the reduced temperature.
As is shown in the upper panel of Fig. \ref{fig:sigmaT_all}(a) for the LSCO ($x$ = 0.10) film and in Fig. 7(b) for the LSCO ($x$ = 0.07) film, $\sigma_2 (T)$ starts to rapidly increase from $T/T_\text{c}^\text{zero} \sim 2$ at most.
In the optimally doped and the overdoped regions, $\sigma_2 (T)$ starts to increase from $T/T_\text{c}^\text{zero} \sim 1.2$ (upper panel in Fig. \ref{fig:sigmaT_all}(b)) and $T/T_\text{c}^\text{zero} \sim 1.5$ (upper panel in Fig. \ref{fig:sigmaT_all}(c)), respectively.
This indicate that the superconducting fluctuation in the optimally-doped region is the smallest, which should have some relationship to the dimensionality of the critical dynamics.
Therefore, we believe that the doping dependence of the THz complex conductivity obtained in this work reveals very essential features of HTSC.

In Fig. \ref{fig:PD}, we show the onset temperature of the superconducting fluctuation observed in this work in the electronic phase diagram of LSCO, which is defined as the starting temperature of the rapid increase of $\sigma_2 (T)$ ($T_\text{fluct}$, open triangles).
The open circles and squares show the corresponding results of the microwave study,\cite{Ohashi2009PRB} where the analytically obtained temperatures were depicted for the underdoped region (circle).
The solid lines are the onset temperature of the Nernst signal (Nernst, gray)\cite{Wang2006PRB} and pseudogap temperature observed by the photoemission experiment (pseudogap, olive).\cite{Hashimoto2007PRB}
$T_\text{fluct}$ is almost equal to the temperatures of the fluctuation onset obtained in the microwave study\cite{Ohashi2009PRB} and a recent report of THz conductivity\cite{Bilbro2011NPhys} but is different from that of the Nernst and diamagnetic signal.
Since we determine $T_\text{fluct}$ from $\sigma_2$, which directly relates to the superfluid density, we believe that the high onset temperature of the Nernst signal in LSCO is not related to the superconducting fluctuation we argued but may rather be related to some types of the charge order, which was observed in Nd- and Eu-doped LSCO\cite{Cyr-Choiniere2009Nature} (described in the introduction, for instance).

\begin{figure*}[!t]
\begin{center}
\includegraphics[width=0.8\linewidth]{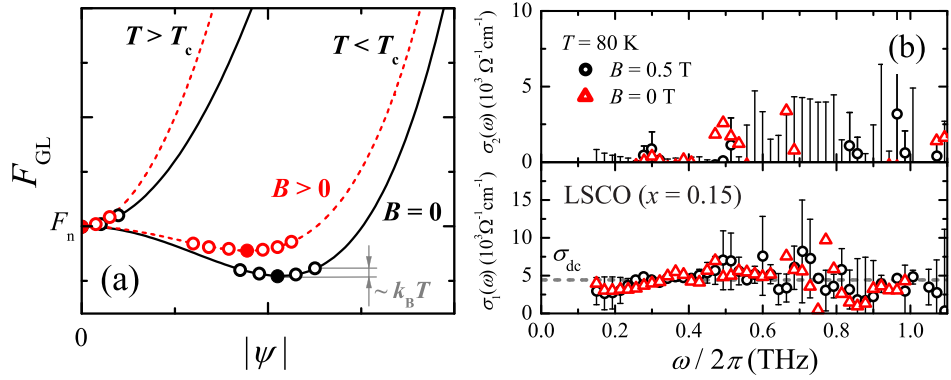}
\caption{\label{fig:mag} (Color online) (a) Schematic figure of the order parameter dependence of the GL free energy, $F_\text{GL} (|\psi|)$. The solid lines are $F_\text{GL}$ under a zero magnetic field, and the broken lines are that under a finite magnetic field. The closed circle indicates the lowest energy position, and the open circles indicate the possible positions by the thermal fluctuation, $k_\text{B}T$.
(b) $\tilde \sigma (\omega)$ of the optimally doped LSCO ($x$ = 0.15) film at 80 K under 0.5 T (open circles) and under 0 T (open triangles). The broken line in the $\sigma_1$ panel is $\sigma_\text{dc}$.}
\end{center}
\end{figure*}

\subsection{Magnetic field effect}

Finally, we discuss the effect of magnetic fields on the superconducting fluctuation.
Within the Ginzburg-Landau (GL) theory, the superconducting fluctuation by the thermal fluctuation energy becomes smaller under a finite magnetic field.
To see this behavior schematically, we depict the dependence of the superconducting order parameter of the GL free energy ($F_\text{GL}$) in Fig. \ref{fig:mag}(a).\cite{Skocpol1975RPP}
$F_\text{GL}$ is described as\cite{TinkhamBook}
\begin{equation}
 F_\text{GL} = F_\text{n} + \alpha |\psi |^2 + \frac{\beta}{2} |\psi |^4 + \frac{1}{2m^\ast} \left| \left(-i\hbar \vec \nabla  -e^\ast \vec{A} \right) \psi \right|^2 + \frac{\vec{B}^2}{2\mu_0 },
 \label{eq:GLenergy}
\end{equation}
where $F_\text{n}$ is the free energy in the normal state without the thermal fluctuation.
In Fig. \ref{fig:mag}(a), solid lines are $F_\text{GL} (|\psi |)$ under zero magnetic field, and broken lines are those under a finite magnetic field.
Closed circles indicate the lowest energy position, and open circles indicate possible positions under the finite thermal fluctuation, $k_\text{B}T$.
With finite thermal fluctuation and under a finite magnetic field, the fluctuation of the superconducting order parameter (open circles) increases at $T < T_\text{c}$ and decreases at $T > T_\text{c}$ compared with that under a zero magnetic field.
However, if the nature of the superconducting fluctuation above $T_\text{c}$ is unusual and enhanced by the magnetic field, we might expect that the superconducting fluctuation survives up to very high temperatures above $T_\text{c}$.
To check this, we investigated the THz conductivity of the optimally doped LSCO ($x$ = 0.15) film under a 0.5 T magnetic field at 80 K and 50 K.
For the experiment under the finite magnetic field, we used a different cryostat with a split-pair type superconducting magnet (Oxford Instruments, Spectromag 4).
Figure \ref{fig:mag}(b) shows $\tilde \sigma (\omega)$ at 80 K under a 0.5 T magnetic field (open circles) and zero magnetic field (open triangles).
The broken line in the $\sigma_1$ panel is $\sigma_\text{dc}$ at 80 K.
We could not find substantial difference between $\tilde \sigma (\omega)$ under 0.5 T and under 0 T within the error-bar.
Thus, in this temperature region, the superconducting fluctuation was not observed in the high-frequency conductivity (this work and the microwave conductivity\cite{Ohashi2009PRB, MaedaTanaka2010PhysC}) whereas a finite Nernst signal and diamagnetic signal do appear.
Therefore, from this result, we insist that the superconducting fluctuation is not enhanced by the magnetic field, at least under 0.5 T.

\section{Conclusion}

We investigated the complex conductivity in the THz region of La$_{2-x}$Sr$_x$CuO$_4$ thin films with various carrier concentrations.
The imaginary part of the complex conductivity starts to increase rapidly from temperatures above $T_\text{c}^\text{zero}$ due to the fluctuating superfluid density with a short lifetime, and the onset temperature of the superconducting fluctuation is sensitive to the carrier concentration.
The superconducting fluctuation starts from $T / T_\text{c}^\text{zero} \sim 2$ for the underdoped sample at most, which is consistent with the conclusion obtained by the microwave conductivity measurement.\cite{Ohashi2009PRB}
The superfluid density is not enhanced under at least a 0.5 T magnetic field.
In addition, we found a notable doping dependence of THz complex conductivity near $T_\text{c}$ and in the superconducting state, which we believe is related to the change in the dimensionality and critical indices of the superconducting fluctuation.
Because our results are different from the results of the Nernst experiment, we believe that the onset temperature of the Nernst signal is not related to the superconducting fluctuation we argued in this paper.

\begin{acknowledgment}

We thank Prof. M. Tonouchi for the photoconductive switch and technical support.
D. Nakamura also thank to the Grant-in-Aid for Scientific Research from Japan Society for the Promotion of Science.
\end{acknowledgment}

\end{document}